\documentclass[runningheads]{llncs}
\usepackage{graphicx}
\usepackage{tabularx}
\usepackage{mathtools}
\usepackage{amsmath}
\usepackage{multirow}
\usepackage{amssymb}

\begin{document} 
\title{Deep Angular Embedding and Feature Correlation Attention for Breast MRI Cancer Analysis}
\titlerunning{Breast MRI Cancer Analysis}
%
\author{Paper ID: 1059}
\author{Luyang Luo\inst{1} \and
Hao Chen\inst{2} \and
Xi Wang\inst{1} \and
Qi Dou\inst{3} \and \\
Huangjing Lin\inst{1} \and
Juan Zhou\inst{4} \and
Gongjie Li\inst{5} \and
Pheng-Ann Heng\inst{1}}

\authorrunning{L. Luo et al.}

\institute{Dept. of Computer Science and Engineering, \\The Chinese University of Hong Kong, Hong Kong SAR, China\and
Imsight Medical Technology, Co., Ltd., China \and
Dept. of Computing, Imperial College London, London, UK \and
Dept. of Radiology, The Fifth Medical Center  \\of Chinese PLA General Hospital, Beijing, China\and
Beijing Image Diagnostic Center of Rimag, Beijing, China
}

\maketitle              

\begin{abstract}
Accurate and automatic analysis of breast MRI plays an important role in early diagnosis and successful treatment planning for breast cancer. Due to the heterogeneity nature, accurate diagnosis of tumors remains a challenging task. In this paper, we propose to identify breast tumor in MRI by Cosine Margin Sigmoid Loss (CMSL) with deep learning (DL) and localize possible cancer lesion by COrrelation Attention Map (COAM) based on the learned features. The CMSL embeds tumor features onto a hyper-sphere and imposes a decision margin through cosine constraints. In this way, the DL model could learn more separable inter-class features and more compact intra-class features in the angular space. Furthermore, we utilize the correlations among feature vectors to generate attention maps that could accurately localize cancer candidates with only image-level label. We build the largest breast cancer dataset involving 10,290 DCE-MRI scan volumes for developing and evaluating the proposed methods. The model driven by CMSL achieved classification accuracy of 0.855 and AUC of 0.902 on the testing set, with sensitivity and specificity of 0.857 and 0.852, respectively, outperforming other competitive methods overall. In addition, the proposed COAM accomplished more accurate localization of the cancer center compared with other state-of-the-art weakly supervised localization method.

\end{abstract}

\section{Introduction}
Breast cancer is the most common malignancy affecting women worldwide \cite{ref_article1}. Early diagnosis of breast cancer is essential for successful treatment planning, where Magnetic Resonance Imaging (MRI) plays a vital role for screening high-risk populations \cite{MRI}. Clinically, radiologists use the Breast Imaging-Reporting and Data System (BI-RADS) to categorize breast lesions into different levels according to their phenotypic characteristics presented in MRI images, indicating different degrees of cancer risk. However, such assessment suffers from inter-observer variance and often subjectively relies on the radiologists' experience. Moreover, due to the heterogeneity nature, tumors of the same pathological result (malignant or benign) could have diverse patterns and hence result in different BI-RADS assessments. In other words, tumors could possess ambiguous inter-class difference and large intra-class variance, which poses a serious challenge to accurate diagnosis of breast cancer. \\

\indent Generally, there are two major tasks regarding to breast MRI tumor analysis: identification of tumors and localization of cancer candidates. Recently, Deep Learning (DL) based approaches have demonstrated great potential in assisting diagnosis of breast cancer in an automatic and fast manner. Previous studies manually annotated tumors and deliberately extracted the corresponding slices or patches for classification \cite{small_lesion,SPIE-2017}. Such methods depended on careful annotations both for training and testing and could not easily be adopted to clinical application. On the other hand,  Guy et al. \cite{unsupervised-MRI} proposed to first automatically localize the lesions and then classify cancer candidates at the second stage. Although the inference in the testing stage thereby was free of lesion delineation, these works still required annotations for model training. To get rid of manual extraction for region of interest (RoI), Gabriel et al. \cite{meta-learn-DCE_MRI} proposed to meta-learn the breast MRI cancer classification problem with only image-level labels. However, all the mentioned studies were limited to small size datasets and consequently lack of generalization validation. More importantly, the relatively low precision or specificity reported in these works implied that the aforementioned problem of inter-class difference and intra-class variance has not been addressed yet.\\

\indent To this end, we propose a Cosine-Margin Sigmoid Loss (CMSL) to tackle the heterogeneity problem for breast tumor classification and COrrelation Attention Map (COAM) for precise cancer candidates localization, both with image-level labels only. The CMSL is extended from the cosine loss originally designed for face verification \cite{cos-face}. It embeds the deep feature vectors onto a hyper-sphere and learns a decision margin between classes in the angular feature space. As a result, the learned features possess more compact intra-class variance and more separable intra-class difference. In addition, we observe a RoI shifting problem of localizing cancer by class activation map \cite{CAM}. Therefore, we propose a novel weakly supervised method, i.e., COAM, to localize cancer candidates more accurately by leveraging deep feature correlations based on the Gram matrix. Furthermore, we build the largest breast DCE-MRI dataset including 10,290 volume scans from 1715 subjects to develop and evaluate our methods.

\begin{figure}[t]
\centering
\includegraphics[width=\textwidth]{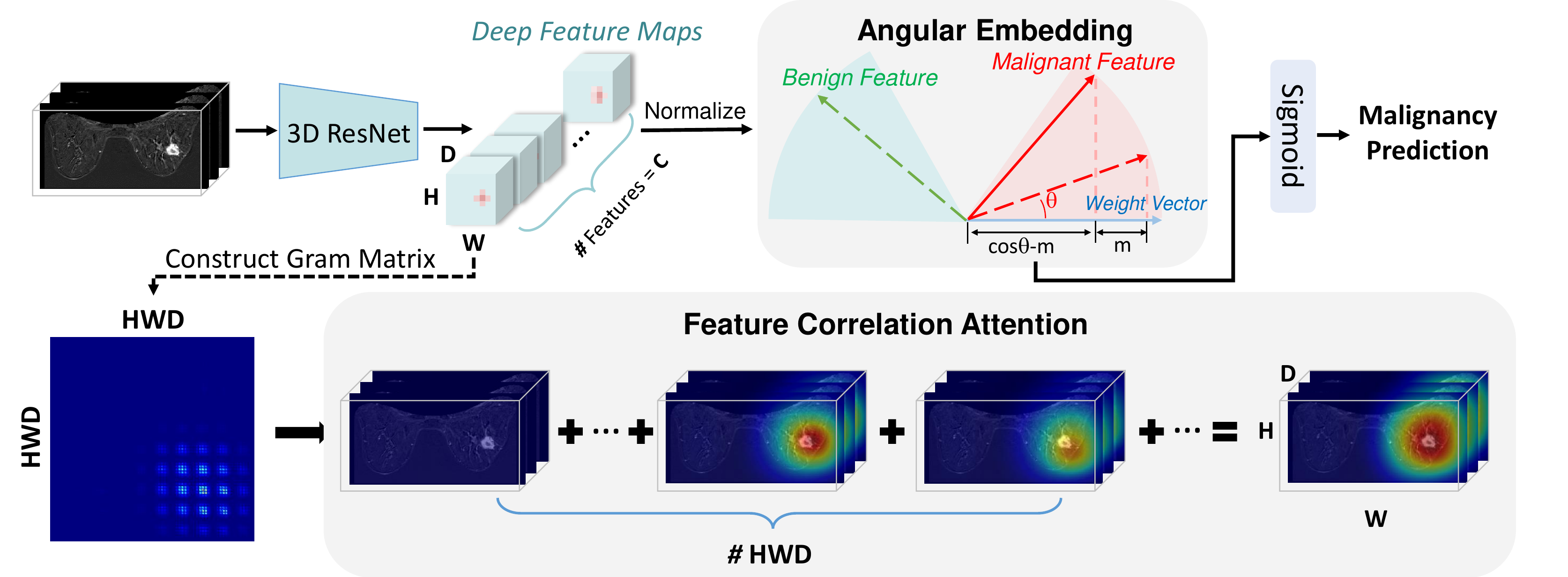}
\setlength{\abovecaptionskip}{-1pt}
\setlength{\belowcaptionskip}{-15pt}
\caption{The framework of breast MRI cancer analysis. 
A 3D ResNet is first trained with CMSL by embedding the deep features onto hyper-sphere. In the testing stage, the deep features are used to construct Gram matrix to obtain correlation attention map.} 
\label{framework}
\end{figure}
 
\section{Methods}
Our framework of breast MRI tumor analysis consists of two parts as illustrated in Fig.~\ref{framework}. One is tumor classification by deep angular embedding driven DL network. The other is weakly supervised cancer candidates localization with feature correlation attention map.

\vspace{-8pt}
\subsection{Cosine Margin Sigmoid Loss for Tumor Classification}
\begin{figure}[t]
\centering
\includegraphics[width=0.8\textwidth]{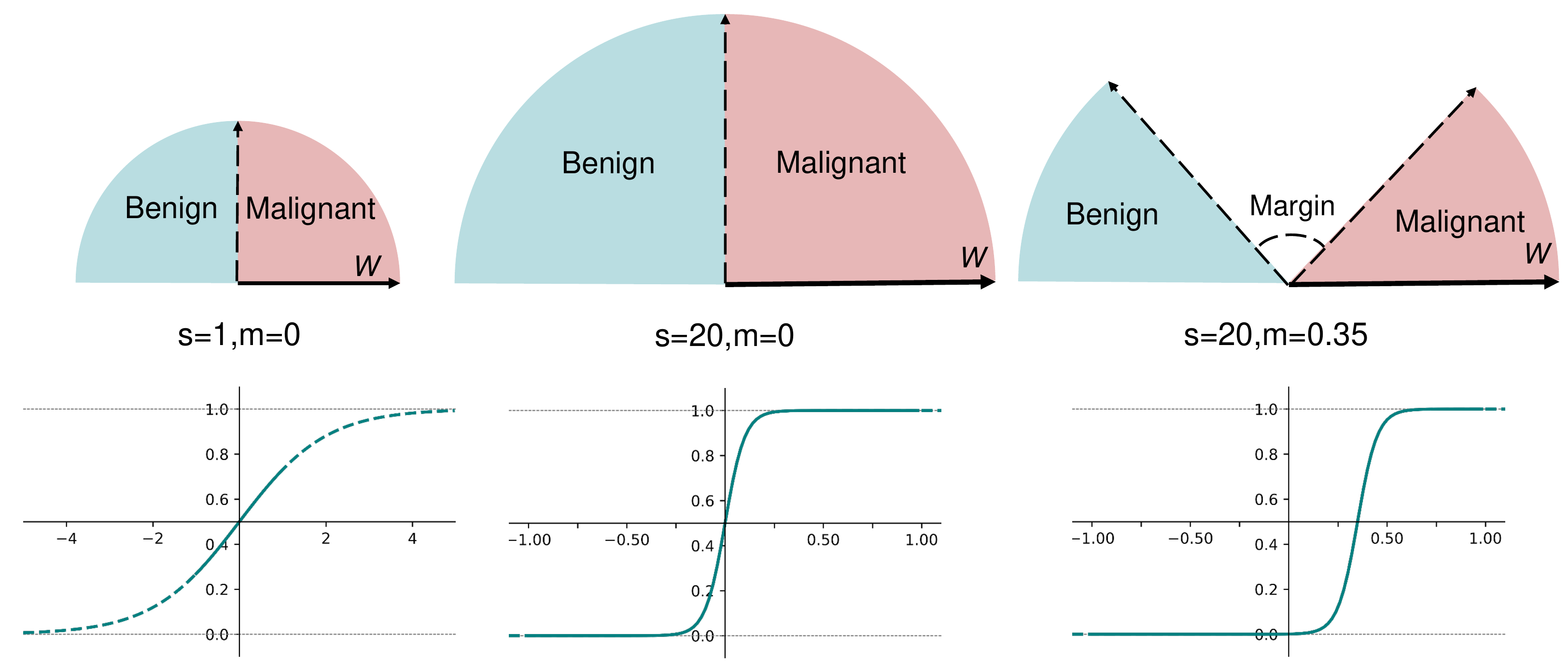}
\setlength{\belowcaptionskip}{-10pt}
\caption{The illustration of NSL with $s=1$, NSL with $s=20$ and CMSL with $s=20$ and $m=0.35$. First row is the geometric interpretation of feature projection on a 2D sphere. Dashed arrows represent the decision boundaries. Second row is the plot of corresponding sigmoid functions. Dashed curves represent the values out of range.} \label{sigmoid}
\end{figure}

The phenotype of tumors has ambiguous inter-class difference and large intra-class variance. Accordingly, the features learned by the DL model could inherit these characteristics. To address this issue, we start by revisiting the traditional sigmoid loss for binary classification problem. Given the input feature vector $x$ of the last fully connected (FC) layer and its corresponding label $y$, the binary sigmoid loss is as follows:

\begin{align}
\mathcal{L}(w;x) =& -y\cdot \text{log}(p(y \mid x))-(1-y)\cdot \text{log}(1-p(y \mid x)) \\
		    =& -y\cdot \text{log}(\frac{1}{1+e^{-w^{T} x}})-(1-y)\cdot \text{log}(1-\frac{1}{1+e^{-w^{T} x}}) 
\end{align}

\noindent where $w$ is the weight parameter of the FC layer, and $p(y \mid x)$ represents the probability of $x$ being classified to $y$. To distinguish different classes, the DL model is expected to give different predictions by adjusting the value of $w^{T}x$. Notice that $w^{T}x=\|w\|\|x\|cos\theta$, where $\theta$ is the angle between feature vector $x$ and weight vector $w$, and $\|\cdot\|$ is the $L_{2}$ norm operation. Generally, the DL model would implicitly alter $\|w\|$ and $\|x\|$ in the Euclidean space and $cos\theta$ in the angular space. However, the aforementioned heterogeneity issue could lead to ambiguous features that are quite hard to discriminate. To this end, constraints on feature distances are considered to regulate the DL model for more separable inter-class features and more compact intra-class features \cite{cos-face}. Since Euclidean distance is not bounded and hence difficult to constrain, we prefer to add regularization on the angular distance which is bounded by $-1 \leq cos\theta \leq 1$. Specifically, we eliminate the influence of the norms $\|x\|$ and $\|w\|$ by modifying the computation of $p(y \mid x)$ to:
\\

\begin{equation} \label{eq:3}
p(y \mid x) = \frac{1}{1+e^{-s\frac{w^{T}x}{\|w\|\|x\|}}} = \frac{1}{1+e^{-s\cdot cos\theta}}
\end{equation}

\noindent where $s$ is a hyper-parameter adjusting the slope of the sigmoid function and controlling the back propagated gradient values. If $s$ is too small, the loss cannot converge to 0 because the sigmoid function is not able to reach its saturation area, given that $-1 \leq cos\theta \leq 1$. On the contrast, if $s$ is set to a large value, the sigmoid function could easily reach the saturation area and result in small gradients, which prevents the network from learning sufficient knowledge. Following \cite{cos-face}, we refer to the loss with modified $p$ in Eq.(\ref{eq:3}) as Normalized Sigmoid Loss (NSL), which focuses on separating features in the angular space with decision boundary $cos\theta=0$ for both classes. Geometrically, we embed the feature vector and the weight vector onto a hyper-sphere whose radius is tuned by $s$. However, the ambiguous features can still distribute near this boundary. Therefore we add an explicit guidance to NSL as follows:

\begin{equation}
\mathcal{L}(w;x) = -y\cdot \text{log}(\frac{1}{1+e^{-s \cdot (cos\theta-I(y) \cdot m)}})-(1-y)\cdot \text{log}(1-\frac{1}{1+e^{-s \cdot (cos\theta-I(y) \cdot m)}})
\end{equation}

\noindent where $I(\cdot)$ is an indicator function. $I(y) = 1$ if $y = 1$ and $I(y) = -1$ otherwise. $m$ is a hyper-parameter that changes the decision boundaries for separating two classes (0 and 1 for benign and malignant) to: $B_{0}: cos\theta + m < 0$ and $B_{1}: cos\theta - m > 0$. Hence a decision margin is imposed by $m$ in the angular space to make the learned inter-class features more separable. Consequently, the distribution space of features shrinks, which eventually leads to more compact intra-class features. Fig.~\ref{sigmoid} shows a comparison among different sigmoid functions and the corresponding geometric illustrations.

\vspace{-8pt}

\subsection{Feature Correlation Attention for Cancer Localization}
Based on the well trained network, localization of cancer candidates can provide more evidences for clinical reference. Therefore, our secondary goal is to localize possible cancers out of other lesion mimics. It is natural for DL studies to use Class Activation Map (CAM) \cite{CAM} for obtaining the Region of Interest (RoI) when only image-level label is available. However, it can not be well generalized to our case due to an observed RoI shifting problem. With the CNN going deeper, the reception fields of neurons become larger, hence neighbors of the tumor feature also capture views over the tumor patch in the image. Since the feature vectors corresponding to different classes could be ambiguous, the classifier layer would possibly tend to find discriminative patterns in the neighbors. Consequently, the corresponding RoI generated by CAM would shift from the desired target.\\

\indent To tackle this problem, we further figure out two insights of our task. First, the feature vectors of the same semantic (malignant or normal) ought to have high correlations with each other. Second, through a series of rectified linear units, the network would implicitly learns large activation values for features related to suspicious cancer patch (with label ``1''), and small activation values for features related to normal patch (with label ``0''). Based on these two intuitions, we leverage the Gram matrix \cite{TextureNIPS} to find the RoI. Given the deep feature map $X \in \mathbb{R}^{H\times W \times S \times C}$ from the last activation layer, where $H, W, S$ and $C$ are the height, width, number of slices and number of channels, respectively, we first reshape $X$ to $X' \in \mathbb{R}^{N \times C}$, where $N=H\times W \times S$. Then we compute an attention vector $M \in \mathbb{R}^{N}$ as follows:
\begin{equation}
M_{i} = \sum_{j=1}^{N}G_{i,j} = \sum_{j=1}^{N}\sum_{k=1}^{C}X'_{i,k}X'_{j,k}
\end{equation}
\noindent where $G \in \mathbb{R}^{N\times N}$ is the Gram matrix over the set of deep feature vectors in $X'$. Each entry $G_{i,j}$ is the inner product of $X'_{i}$ and $X'_{j}$, representing the correlation between $i$-th and $j$-th vector. Because our network is trained for binary classification, it enables the gap between large and small activation values of feature vector related to suspicious cancer and normal patch. Correspondingly, the correlation value would also be relatively large or small according to the activation values of the features. Inspired by \cite{DualAttention}, each column $G_{i}$ can be interpreted as a sub-attention map implying the network's attention of the class that $i$-th vector belongs to. Thus the above operation is equal to element-wise summation over all sub-attention maps $G_{i}$. Moreover, since $G$ is symmetric, the element-wise summation is also equivalent to summing over $G_{i}$ to be the value of $M_{i}$.  Essentially, $\sum_{j=1}^{N} G_{i,j}$ indicates the \textit{importance} of $i$-th feature determined by the sub-attention of the feature map at its $i$-th position. At last, by simply reshape $M$ to $H\times W\times S$ we are able to obtain an attention map purely based on the deep feature correlations. We refer to this method as COrrelation Attention Map (COAM). It is worth mentioning that COAM is related to the self-attention mechanism \cite{DualAttention} and the stationary feature space representation \cite{TextureNIPS}. However, it differs from these works because the Gram matrix is not involved at any optimization stage and directly used for attention map generation. 

\vspace{-10pt}
\section{Experiments and Results}
\vspace{-5pt}
\subsection{Implementation Details}
\subsubsection{Dataset}
We built the largest breast tumor Dynamic Contrast Enhanced (DCE) MRI dataset involving 10,290 scans from 1715 subjects, with 1137 cases containing malignant tumors and 578 cases containing benign tumors. All of the scans were conducted with a 1.5-T Siemens system. We collected 6 DCE-MRI subtraction scans and 1 non-fat suppressed T1 scan from each subject. BI-RADS categories were assessed by 3 radiologists. Pathological labels was given by biopsy or surgery diagnosis. The data were randomly divided into training, validation and testing sets with 1204, 165 and 346 subjects, respectively.

\vspace{-12pt}
\subsubsection{Preprocessing}
Frangi's approach\cite{Frangi Filter} was first applied on the slices of each non-fat suppressed T1 scan to detect evident edges. Next, thresholding, small connected component removal and hole filling were sequentially employed to obtain coarse breast region masks. Afterwards, the 2D masks were stacked into volumes, followed by Gaussian smooth. We used the 3D masks to segment the subtractions. Note that the DCE-MRI and non-fat suppressed scans were originally registered in the scanning machine. Finally we clipped and normalized the intensity values, concatenated 6 subtractions, and cropped or padded the data to a fixed size of $340\times220\times128$ as the model inputs.

\vspace{-12pt}
\subsubsection{Training Strategy}

We used 3D ResNet34 \cite{Resnet} as the base model and replaced the global average pooling layer and FC layer with an $1\times 1\times 1$ convolutional layer appended with a pooling layer. The hyper-parameter $s$ and $m$ were set to 20 and 0.35, respectively, similar to \cite{cos-face}. The learning rate was initially set to $10^{-4}$ and decreased 10 times when training error stagnated. The base model is trained until convergence and then employed to initialize all other methods. 
\begin{table}[b]
\caption{Comparison of different methods on cancer classification.}\label{quantitative table}
\centering
\begin{tabular}{ccccccc}
\hline \hline
                          & Method      & Accuracy       & Sensitivity    & Specificity    & F1             & AUC            \\
\hline

\multirow{4}{*}  & 2D MIL \cite{MIL-cvpr}     & 0.789          & 0.870          & 0.626          & 0.846          & 0.842          \\
                          & 3D ResNet \cite{Resnet}       & 0.821          & 0.840          & 0.783          & 0.862          & 0.880          \\

                         & 3D Sparse MIL \cite{sparse-mil-mammography}      & 0.832          & 0.857          & 0.783          & 0.872         & 0.885 \\
                          & 3D DK-MT \cite{mt-ultrasound} & 0.824          & \textbf{0.896}          & 0.643          & 0.864          & 0.883          \\ 

\multirow{2}{*}  & 3D ResNet+NSL         & 0.821          & 0.840          & 0.783          & 0.862          & 0.874          \\ 
                      & \textbf{3D ResNet+CMSL (ours)}    & \textbf{0.855} & 0.857          & \textbf{0.852} & \textbf{0.888} & \textbf{0.902} \\ \hline \hline
\end{tabular}
\vspace{-5pt}
\end{table}
\vspace{-10pt}
\subsection{Evaluation and Comparison}
\subsubsection{Tumor Classification}
We conducted comparison among several deep learning methods: (1)\textit{2D MIL}: a multi instance method aggregating features from 2D slices by 2D ResNet34 \cite{MIL-cvpr}; (2)\textit{3D ResNet}: a 3D implementation of ResNet34; (3)\textit{3D Sparse MIL}: a sparse label assign method \cite{sparse-mil-mammography}; (4)\textit{3D DK-MT}: a domain knowledge driven multi-task learning network \cite{mt-ultrasound}; (5)\textit{3D ResNet+NSL}: Normalized sigmoid loss based on (2); (6)\textit{3D ResNet+CMSL}: our proposed CMSL based on (2). We computed the accuracy, specificity, sensitivity, F1 score and AUC as the evaluation metrics. Experimental results are reported in Table \ref{quantitative table}.\\
\indent Compared with 2D method, 3D approaches achieved better results by utilizing more spatial information. Both \textit{3D Sparse MIL} and \textit{3D DK-MT} adopted additional assumption or knowledge, leading to better performance than vanilla \textit{3D ResNet}. Noticeably, \textit{3D DK-MT} showed poor specificity, which is possibly due to imbalanced auxiliary knowledge (more BI-RADS 4 and 5 than 3) that dominated the learning process. For deep angular embedding based methods like \textit{3D ResNet+NSL}, simply taking the features into angular space without margin constraint caused certain performance decay. It implied that the network cannot learn sufficient knowledge if $s$ is set to a large value. Moreover, our proposed \textit{3D ResNet+CMSL} method significantly improved the results. The underlying reason is that it could learn more discriminative patterns by imposing cosine margin. Our method achieved the highest specificity with over 7.9\% better than all other methods and kept a comparable sensitivity at the mean time. It exceeded all other methods with over 2\% in AUC, over 3\%  in accuracy and over 1.5\% in F1 score, proving that addressing the inter- and intra-class problem can improve performance of breast tumor classification.

\begin{figure}[t]
\centering
\includegraphics[width=0.8\textwidth]{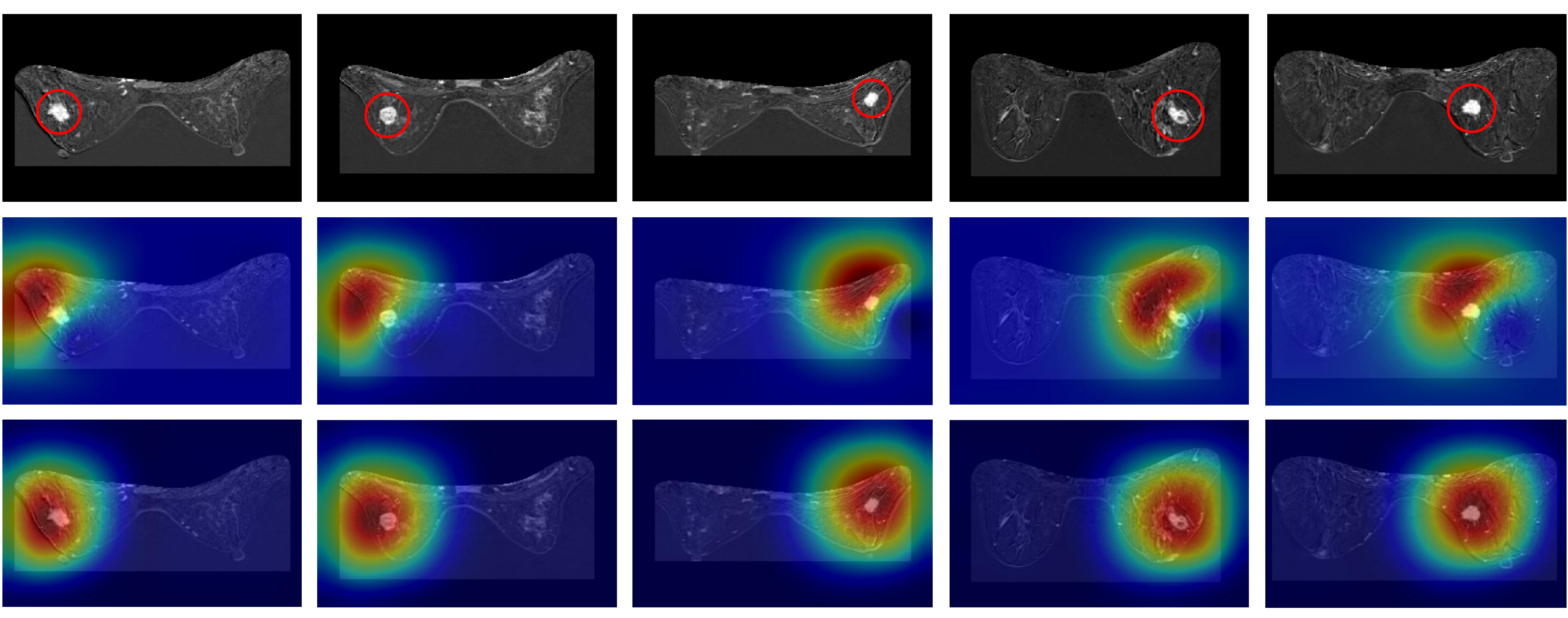}
\setlength{\belowcaptionskip}{-10pt}
\caption{Comparison Between CAM and COAM. We select typical slices from different subjects for a qualitative demonstration. First row: DCE-MRI subtraction slice; second row: visualization of CAM; third row: visualization of COAM. Cancer lesions are circles by red. Best viewed in color.} 
\label{Heatmaps}
\end{figure}

\vspace{-10pt}
\subsubsection{Cancer Localization}
To evaluate the performance of COAM, we invited the radiologists to manually annotate 85 samples that were classified as malignant by our model. We compared our method with CAM by computing the Euclidean distance between center position of the annotation and the voxel position with highest value in the attention map. Then the distance is multiplied by the voxel spacing, i.e., 1.1 mm, as the final measurement. The criteria is reported in the form of $mean\pm std$, where $mean$ and $stdv$ represent the mean value and standard deviation of the center distances over 85 samples, respectively. Compared to the distance of 39.84$\pm$8.82mm by CAM, COAM showed a significant advantage with 18.26$\pm$13.65 mm only. Fig.~\ref{Heatmaps} showed the qualitative comparison with these two methods.
\vspace{-10pt}

\section{Conclusion}
In this paper, we propose the cosine margin sigmoid loss for breast tumor classification and correlation attention map for weakly supervised cancer candidates localization based on MRI scans. First, we use CMSL driven deep network to learn more separable inter-class features and more compact intra-class features which effectively tackle the heterogeneity problem of tumors. In addition, the proposed COAM leverages correlations among deep features to localize region of interests in a weakly supervised manner. Extensive experiments on our large-scale dataset demonstrates the efficacy of our methods which outperform other state-of-the-art approaches significantly on both tasks. Our methods are general and can be extended to many other fields.

\end{document}